\documentclass[twocolumn,showpacs,showkeys,superscriptaddress]{revtex4}

\usepackage{amsmath}
\usepackage{bbold}
\usepackage{amsfonts}
\usepackage{amssymb}
\usepackage{pbsi}
\usepackage[T1]{fontenc}
\usepackage{hyperref}
\usepackage{xcolor}
\usepackage{graphicx}
\usepackage{subfig}
\usepackage{float}

\def\openone{\leavevmode\hbox{\small1\kern-3.8pt\normalsize1}}
\def\N{\leavevmode\hbox{ Z \kern-8 pt\normalsize{Z}}}
\def\openone{\leavevmode\hbox{\small1\kern-3.8pt\normalsize1}}
\def\openJ{\leavevmode\hbox{J \kern-9.5pt\normalsize J}}
\def\openS{\leavevmode\hbox{ S \kern-9.3pt\normalsize S}}

\newcommand{\bb}{\begin{equation}}
\newcommand{\ee}{\end{equation}}
\newcommand{\eqb}{\begin{eqnarray}}
\newcommand{\eqf}{\end{eqnarray}}

\usepackage{color}

\begin{document}

\title{Supersymmetric relativistic quantum mechanics in time-domain}

\author{Felipe A. Asenjo}
\email{felipe.asenjo@uai.cl}
\affiliation{Facultad de Ingenier\'ia y Ciencias,
Universidad Adolfo Ib\'a\~nez, Santiago 7491169, Chile.}
\author{Sergio A. Hojman}
\email{sergio.hojman@uai.cl}
\affiliation{Departamento de Ciencias, Facultad de Artes Liberales,
Universidad Adolfo Ib\'a\~nez, Santiago 7491169, Chile.}
\affiliation{Departamento de F\'{\i}sica, Facultad de Ciencias, Universidad de Chile,
Santiago 7800003, Chile.}
\affiliation{Centro de Recursos Educativos Avanzados, CREA, Santiago 7500018, Chile.}
\author{H\'ector M. Moya-Cessa}
\email{hmmc@inaoep.mx}
\affiliation{Instituto Nacional de Astrof\'isica, \'Optica y Electr\'onica. Calle Luis Enrique Erro No. 1,
Santa Mar\'ia Tonantzintla, Puebla 72840, Mexico.}
\author{Francisco Soto-Eguibar}
\email{feguibar@inaoep.mx}\affiliation{Instituto Nacional de Astrof\'isica, \'Optica y Electr\'onica. Calle Luis Enrique Erro No. 1,
Santa Mar\'ia Tonantzintla, Puebla 72840, Mexico.}

\begin{abstract}
A  supersymmetric relativistic quantum theory in the temporal domain is developed for bi-spinor fields satisfying the Dirac equation. The simplest time-domain supersymmetric theory can be postulated
for  fields with time-dependent mass, showing an equivalence with the bosonic supersymmetric theory in time-domain. Solutions are presented and they are used to produce probability oscillations between   mass states. 
As an application of this idea, we study the two-neutrino oscillation problem, showing that  flavour state oscillations may emerge from the supersymmetry originated by
the time-dependence of the unique mass of the neutrino. 
\end{abstract}


\maketitle

\section{Introduction}
Supersymmetry is one of the cornerstones of theoretical physics \cite{freund}, being ubiquitous  in almost every branch of physics \cite{negro,David,coop1,coop2,crom,jfeng,hall,zach,haber,aiit,qcosmo1,qcosmo2,qcosmo3,hojmanasenjo,carena,simon,cmscol,vladi,rouz,heweet,sunyu,Plyushchay1,Plyushchay2,correaf,Efetov,junker,Correa22,mateosG,nieto22,Plyushchay3,adrianS,fazsah17,carlos}. The standard way to proceed is to construct supersymmetric theories in a space-domain, where the superpartners and supercharge operators take into account the spatial variations of, for example, external potentials. In this work, it is not our aim to focus in those spatial supersymmetries, but instead to inquire if a temporal version of such theories is possible for massive fields in relativistic quantum mechanics.

Recently,  in Refs.~\cite{fazsah17,carlos}, it was introduced the concept of supersymmetry in time-domain (T-SUSY) for Maxwell equations. This is a  supersymmetry 
occurring in the temporal part of the massless field dynamics, completely uncoupled from the spatial evolution of the field. They studied the applications of T-SUSY in the realm of optics for dispersive media, showing the novel capabilities of this theory to introduce hypothetical  materials with new optical features. Besides, they showed that  this time-domain supersymmetry applies to any field described, in principle, by a d'Alambertian equation. In other words, they developed the bosonic version of the T-SUSY theory. 

Supersymmetric theories for purely bosonic systems have been extensively developed \cite{Plyushchay1,correaf, Plyushchay2}. On the contrary, it is the purpose of this manuscript to show that a T-SUSY theory  can also be obtained in relativistic quantum mechanics for fields described by the Dirac equation, thus complementing the usual spatial supersymmetry  theory \cite{spinspatialsusy}. In this case, we show that the simplest T-SUSY theory may be constructed for  time-dependent massive fields satisfying Dirac equation, finding  solutions for different possible time-dependent masses. Also, this theory is  equivalent to its bosonic partner, allowing us to obtain a massive particle field behavior that is analogue to a light-like one.

Besides, this theory produces probability states oscillations as a consequence of its supersymmetry. Thus, it can be used to study the neutrino oscillation problem. In this way, we give a different perspective to the origin of  neutrino oscillations through supersymmetry in time-domain.  

\section{T-SUSY for Dirac equation}
Let us consider a bi-spinor field $\Psi$ satisfying the Dirac equation in flat spacetime for a field with mass $m$,
\begin{equation}
    i\gamma^\mu{\partial_\mu} \Psi=m\Psi\, ,
    \label{Dirac10}
\end{equation}
where  
$\partial_\mu$ are the covariant derivatives. Here, $\gamma^\mu$ are the
gamma matrices in the Dirac representation, fulfilling 
$\gamma^\mu\gamma^\nu+\gamma^\nu\gamma^\mu=2\eta^{\mu\nu}\mathbb{1}_{4\times4}$,
with the flat spacetime metric  $\eta^{\mu\nu}={\mbox{diag}}(1,-1,-1,-1)$, and
 with the   identity matrix $\mathbb{1}_{4\times4}$.  The cases with external potential are discussed in the last section, but for now it is enough to consider the T-SUSY theory for Eq.~\eqref{Dirac10}. 
 
 In the following construction for a relativistic quantum mechanical T-SUSY, from Eq.~\eqref{Dirac10}, a time-dependent mass is essential.  In principle, a time-dependent mass can be understood as an interacting Lorentz scalar potential field \cite{coop2}.  It is important to remark that  conservation of probability associated to bi-spinor fields
 only require real mass, and not fields with constant masses.
 
Let us consider the following form for the bi-spinor
\begin{equation}
    \Psi=\left(\begin{array}{c}
       \Psi_+ \\
        \Psi_- 
    \end{array}\right)\, ,
    \label{wavefunccompoent}
\end{equation}
in terms of spinors $\Psi_\pm$. 
Thereby,  Eq.~\eqref{Dirac10} becomes 
\begin{eqnarray}
   \left( i\frac{\partial}{\partial t} \mp m \right) \Psi_\pm=- i \sigma^j \partial_j\Psi_\mp \, .
\label{eqmecsusyG2noW}
\end{eqnarray}
In order to bring the usual aspects of supersymmetric theories, let us perform a Wick rotation in time and space, $t \rightarrow i t$ and ${\bf x} \rightarrow i {\bf x}$. This is equivalent to a change in the flat spacetime metric signature.  Under this change,  Eq.~\eqref{eqmecsusyG2noW} becomes 
\begin{eqnarray}
    Q_\pm \Psi_\pm=-  \sigma^j \partial_j\Psi_\mp \, ,
\label{eqmecsusyG2}
\end{eqnarray}
with operators
\begin{eqnarray}
    Q_\pm=\frac{\partial}{\partial t} \mp m\, .
\end{eqnarray}
 Now, let us consider a separable spinor of the form
\begin{equation}
  \Psi_\pm(t,x^j)=\psi_\pm(t)\, \chi_\pm(x^j)\, ,
\label{ansatzspinor0}
\end{equation}
with time-dependent functions $\psi_\pm$, and space-dependent spinor $\chi_\pm$.  Using \eqref{ansatzspinor0} in Eq.~\eqref{eqmecsusyG2}, we obtain
\begin{equation}
    \left(Q_\pm \psi_\pm\right) \chi_\pm=-  \psi_\mp \sigma^j \partial_j\chi_\mp \, ,
\label{eqmecsusyG2b}
\end{equation}
where $Q_\pm$  operates on the function $\psi_\pm$.
In this way, the T-SUSY formalism can be achieved for the functions $\psi_\pm$ as they fulfill
\begin{equation}
    Q_\pm \psi_\pm=-  \left(\frac{\chi_\pm^\dag\sigma^j \partial_j\chi_\mp}{\chi_\pm^\dag \chi_\pm}\right)\psi_\mp \, .
\label{eqmecsusyG}
\end{equation}
Although this equation is general, along this work and in order to extract the main physical information from the simplest model, we study the case  $\chi_\pm^\dag\sigma^j \partial_j\chi_\mp=- k\, \chi_\pm^\dag \chi_\pm$ for an arbitrary constant  $k$. In the Dirac basis, this can be easily solved by the following ansatz  for the spinor
\begin{equation}
  \chi_\pm(x^j)= e_\pm  \exp\left(-k_1 x^1\mp i k_2 x^2\right) f(x^3)\, ,
\label{ansatzspinor}
\end{equation}
in terms of an arbitrary function $f$, with $k=k_1+k_2$,
and spinors
\begin{equation}
    e_+=\left(\begin{array}{c}
       1 \\
        0 
    \end{array}\right)\, ,\quad 
    e_-=\left(\begin{array}{c}
       0 \\
        1 
    \end{array}\right),
\end{equation}
with the properties $e_\pm^\dag e_\pm=1$, and $e_\pm^\dag e_\mp=0$. This allows us to reduce the system \eqref{eqmecsusyG} into the form
\begin{eqnarray}
    Q_\pm \psi_\pm= k\, \psi_\mp \, ,
\label{eqmecsusy}
\end{eqnarray}
where $k$ can be shown to play the role of the momentum of the particle.
Eqs.~\eqref{eqmecsusy} correspond to a set of supersymmetric equations for  quantum mechanics,  obtained   in the Dirac matrices basis. 
It is the simplest form of a T-SUSY theory in relativistic quantum mechanics,    and it has an equivalent form to  supersymmetric theory in space-domain \cite{coop2}.

This form of the T-SUSY theory requires that the mass be responsible for the origin of the superpotential of this supersymmetry. This occurs when the mass becomes  time-dependent,  $m=m(t)$, and it can be seen using Eqs.~\eqref{eqmecsusy} to calculate the equations for each function $\psi_\pm$,
\begin{equation}
    H_\pm \psi_\pm=k^2 \psi_\pm\, ,
    \label{eq7hamiltonian}
\end{equation}
where we have defined the super-partners Hamiltonians $H_\pm$ in terms of superpotentials $W_\pm$ as
\begin{eqnarray}
H_\pm=Q_\mp Q_\pm=\frac{d^2}{dt^2}+W_\pm\, ,
\end{eqnarray}
where
\begin{equation}
    W_\pm=
\mp \frac{d m}{dt}-m^2\, ,
\end{equation}
such that $W_- -W_+=2\,  dm/dt$.
Thereby, the difference between states lies in the time dependence of the mass. 

In general, from Eq.~\eqref{eq7hamiltonian}, we can write the functions
\eqref{wavefunccompoent} as
\begin{equation}
    \psi_\pm(t)=\exp\left( -\int dt\, E_\pm\right)\, ,
    \label{wavefuntionconE}
\end{equation}
where $E_\pm$ are   time-dependent functions that satisfy the Ricatti equation
 \begin{equation}
\frac{d E_\pm}{dt}-E_\pm^2 +k^2=W_\pm\, ,
\label{ecuacionparazeta}
\end{equation}
such that $\psi_\pm(0)=1$; whenever $dm/dt\neq 0$, then $E_+\neq E_-$. In this way, for a specific time functionality of the mass, solutions for $E_\pm$ can be found in order to satisfy this T-SUSY description. On the contrary, for constant mass, we obtain the solution $E_\pm^2={k^2+m^2}$ (from where we identify $E$ and $k$ as the particle energy and momentum, respectively) and the T-SUSY aspects of the theory vanish. 

All the framework for a standard supersymmetric theory in quantum mechanics can be  utilized for this T-SUSY theory. For instance, its algebra is defined by the super-Hamiltonian matrix operator ${\bf H}$, and the super-charge matrix operators ${\bf Q}$, defined as
\begin{equation}
    {\bf H}=\left(\begin{array}{cc}
       H_+ & 0 \\
        0 & H_-
    \end{array}\right)\, , 
     {\bf Q}=\left(\begin{array}{cc}
       0 & 0 \\
        Q_+ & 0
    \end{array}\right)\, , 
     {\bf Q}^\dag=\left(\begin{array}{cc}
       0 & Q_- \\
        0 & 0
    \end{array}\right)\, , 
   \end{equation}
and that have the closed algebra ${\bf H}=\{{\bf Q},{\bf Q}^\dag\}$, $[{\bf H},{\bf Q}]=0=[{\bf H},{\bf Q}^\dag]$, $\{{\bf Q},{\bf Q}\}=0=\{{\bf Q}^\dag,{\bf Q}^\dag\}$, for the bosonic ${\bf H}$ and fermionic ${\bf Q}$ and ${\bf Q}^\dag$ operators.  In the same way, as $\psi_\pm$ are the eigenfunctions of $H_\pm$, the spectrum of the two Hamiltonians is thus
degenerate, except that one of the Hamiltonians has an extra state at zero energy. Therefore, using Eqs.~\eqref{eqmecsusy}, a partner eigenfunction can be converted into the other one with the same energy, creating or destroying  an extra node in that eigenfunction \cite{coop2}.

The non-relativistic limit of the above theory can be obtained directly from Eq.~\eqref{eqmecsusyG2noW}; by applying the $(i\partial_t \pm m)$ operator, we obtain $(-\partial_t^2\mp i \partial_t m-m^2)\Psi_\pm=-\sigma^j\sigma^k\partial_j\partial_k\Psi_\pm=-\nabla^2\Psi_\pm$. The non-relativistic limit of this T-SUSY theory is obtained when $\Psi_\pm= \zeta_\pm \exp(\mp i\int m\,  dt)$, 
and
the eikonal approximation is taken $\partial_t^2\zeta_\pm\ll m\,  \partial_t\zeta_\pm$. In such case, we get the usual free-particle Schr\"odinger equation $\pm 2 i m \, \partial_t\zeta_\pm+\nabla^2\zeta_\pm=0$.

 It is  worth to remark that non-relativistic versions of spatial supersymmetry can be constructed to deal with time-dependent Hamiltonians \cite{Vladislavg,Vladislavg2,Vladislavg3,juliacen}. This is still a spatial SUSY theory, but considering the dynamics of time-dependent systems. This is different to what has been developed above for relativistic  time-domain SUSY, or what was first described in Refs.~\cite{fazsah17,carlos}.

Finally, on the other hand, we notice that this relativistic quantum mechanical T-SUSY theory can be put in analogue fashion to its bosonic (Klein-Gordon) counterpart developed in Ref.~\cite{carlos}. We can   rewrite Eq.~\eqref{eq7hamiltonian}  in the form
\begin{equation}\label{0100}
    \left(\frac{ d^2}{dt^2} -\frac{1}{n^2_\pm}\right)\psi_\pm=0\, ,
\end{equation}
where 
\begin{equation}
n_\pm(t)=\left(k^2-W_\pm\right)^{-1/2}\, .
\end{equation}
Eq.~\eqref{0100} is equivalent to the one  describing the propagation
 of light in a medium with supersymmetric
refraction indices $n_\pm$. In this T-SUSY theory, both refraction indices are related by
\begin{equation}
   \frac{1}{n_+^2}={\frac{1}{n_-^2}+2\frac{ d m}{dt}}\, .
\end{equation}
This relation shows the equivalence of fermionic T-SUSY with bosonic T-SUSY   discussed in  Ref.~\cite{carlos}.

\section{Simple solutions}
We can  get simple solutions  for particle  fields with time-dependent mass by first considering   $k\lll m(t)$. For this case, from Eq.~\eqref{ecuacionparazeta}, we find
\begin{equation}
    E_\pm(t)\approx \mp m(t)\, .
\end{equation}
Therefore, the two mass states have different behavior stemming from just one time-dependent mass.

Similarly, another simple solution can be obtained in the light-mass (ultra-relativistic) case, when $k\gg m(t)$. For this case, $k$ corresponds (approximately) to the energy of the particle.
In this case, from Eq.~\eqref{ecuacionparazeta}, we can find an approximated form for $E_\pm$ given by
\begin{equation}
E_\pm(t)\approx k+ e^{2k t}\int dt\,  W_\pm\,  e^{-2kt}\, .
\label{ultrarelativis0}
\end{equation}

\section{Oscillations in T-SUSY}

The above T-SUSY theory allows now to consider the phenomenon of oscillation  
of states in a different fashion. These {\it T-SUSY oscillations} have their origin in the subjacent supersymmetry due to the temporal dependence of the mass.   In order to obtain these physical states, let us return to the physical time $t$ from Eq.~\eqref{Dirac10}, through an inverse Wick rotation applied to the previous solutions.

Considering the bi-spinor \eqref{wavefunccompoent}, let us now define a bi-spinor with mixed states 
\begin{equation}
    \Phi=\left(\begin{array}{c}
       \Phi_a \\
        \Phi_b 
    \end{array}\right)=\left(\begin{array}{cc}
       \cos\theta\, \mathbb{1}_{2\times2} & \sin\theta\, \mathbb{1}_{2\times2} \\
        -\sin\theta\, \mathbb{1}_{2\times2} & \cos\theta\, \mathbb{1}_{2\times2}
    \end{array}\right)\Psi\, ,
    \label{wavefunccompoentmixed}
\end{equation}
with the spinor wavefunctions $\Phi_a$ and $\Phi_b$ defining new states,   the mixing angle in vacuum $\theta$ \cite{kayser,mondal,fantini,petcov}, and  the  $2\times 2$ identity matrix $\mathbb{1}_{2\times2}$.

The amplitude of mixed states change, from $\Phi_a(t=0)$ to $\Phi_b(t)$, can be calculated to be
${\mbox{Amp}}\left(\Phi_a\rightarrow\Phi_b\right)=\sin 2\theta\left[\psi_+^*(0)\psi_+(t)-\psi_-^*(0)\psi_-(t)\right]/{2}$.
Finally, the probability of mixed states change is given by
\begin{align}
    {{P}}\left(\Phi_a\rightarrow\Phi_b\right)=&|{\mbox{Amp}}\left(\Phi_a\rightarrow\Phi_b\right)|^2\nonumber\\
    =&{\sin^2 2\theta}\,  e^{-\alpha}\left( \sinh^2\beta+\sin^2\rho\right)\, ,
    \label{probabilityGen}
\end{align}
where
\begin{align}
    \alpha=&\int dt\left({\mbox{Im}}\left\{E_-\right\}+{\mbox{Im}}\left\{E_+\right\} \right)\, ,\nonumber\\
    \beta=&\frac{1}{2}\int dt\left({\mbox{Im}}\left\{E_-\right\}-{\mbox{Im}}\left\{E_+\right\} \right)\, ,\nonumber\\
    \rho=&\frac{1}{2}\int dt\left({\mbox{Re}}\left\{E_-\right\}-{\mbox{Re}}\left\{E_+\right\} \right)\, .
\end{align}
The terms proportional to $\exp(-\alpha)$ and $\sinh^2\beta$ enter in the probability due to the contribution of the possible time-dependent  amplitude of functions \eqref{wavefuntionconE}. Their phases only contribute to the  probability through $\sin^2\rho$.

In this way, the T-SUSY oscillations between the mixed states, given by the above transition probability, are due only to the different solutions 
of Eq.~\eqref{ecuacionparazeta}. This  occurs if the  field has a non-constant mass, with the T-SUSY properties.

\subsection{Oscillations for massive fields}
As an example, let us evaluate the above probability \eqref{probabilityGen} with the previous simple example for the solution of a massive field, namely  $k\lll m$, and $E_\pm\approx\mp m$. For this case, ${\mbox{Im}}\{E_\pm\}\approx 0$, 
and $\alpha=0=\beta$; then the probability \eqref{probabilityGen} reduces to
\begin{equation}
    P\left(\Phi_a\rightarrow\Phi_b\right)\approx \sin^2 2\theta\, \sin^2\left(\int dt\, m\right)\, .
    \label{probabilityGen2}
\end{equation}
Thus, the T-SUSY oscillations for the massive case occurs because the mass can evolve in time. On the contrary, with constant mass, then $E_+=E_-$ and no oscillation can occur.

\subsection{Oscillations for ultra-relativistic fields}
Let us model the particle oscillation considering the previous simple solution \eqref{ultrarelativis0} for an ultra-relativistic particle, with $m\ll k$. 
We consider the simplest form $m(t)=m_0 \sin(\Lambda t)$ for a time-varying mass, where constant $m_0\ll k$, and $\Lambda$ is a  constant measuring the time-scale variation of mass. In this way, in this model, the particle has a mass that oscillate between a null and a finite value, with temporal span of
$0\leq t\leq \pi/\Lambda$ (in order to avoid negative masses).
Consequently, in this case, solution \eqref{ultrarelativis0} gives
\begin{align}
    E_\pm & \approx k + \frac{m_0^2}{4k} \pm 
    \frac{ m_0\Lambda^2}{4k^2-\Lambda^2}\sin(\Lambda t)
    -\frac{k m_0^2}{4(k^2-\Lambda^2)}\cos(2\Lambda t)
    \nonumber\\ &
    \pm i \frac{2 km_0\Lambda}{4k^2-\Lambda^2}
    \cos(\Lambda t)
    +i\frac{m_0^2\Lambda}{4(k^2-\Lambda^2)}
    \sin(2\Lambda t).
\end{align}
In this way, the total probability for flavour states change can be calculated from Eq.~\eqref{probabilityGen} to be
\begin{eqnarray}
    P\left(\nu_e\rightarrow \nu_\mu\right)&=& \sin^2 2\theta\,\exp\left(\frac{m_0^2}{4(\Lambda^2-k^2)}(\cos(2\Lambda t)-1) \right)\nonumber\\
    &&\times\left[\sinh^2\left(\frac{2 k m_0}{4k^2-\Lambda^2}\sin(\Lambda t)\right)\right.
    \nonumber\\&&
    \left.\quad +\sin^2\left(\frac{m_0 \Lambda}{4k^2-\Lambda^2}\left[\cos(\Lambda  t)-1\right]\right)\right]\, .
    \label{probabilityGenNeutrino0}
\end{eqnarray}

\subsection{Application to the two-neutrino oscillation problem}

We now invite the reader to re-think the two-neutrino oscillation problem in terms of the T-SUSY theory developed above.
Neutrino oscillations have been studied in the context of supersymmetry  \cite{ma,romao,owong,owong2, radovan,valle,faseler,vergados,fengli,arkani,rati,gosh,takaya}, but not within the context of the time-domain theory (with  time-dependent mass fields).

In this way, the  electron and muon flavour  eigenstates can now be identified by 
the bi-spinor \eqref{wavefunccompoentmixed}, while the oscillation probability can be calculated from Eq.~\eqref{probabilityGenNeutrino0} under the following assumptions.
Let us consider  MeV
neutrinos ($k\sim 10^6$ eV), with an upper mass limit of the order 
$m_0\sim 10^{-1}$ eV \cite{KATRIN,Olive},
and    oscillation
distances of tenth of kilometers \cite{Olive}.
Let us assume that $\Lambda\ll m_0\ll k$, in order that the arguments of $\sinh$ and $\sin$ in probability \eqref{probabilityGenNeutrino0} are small. 
Thereby, in this case, the probability \eqref{probabilityGenNeutrino0} for the flavour states change reduces  simply to
\begin{eqnarray}
   \frac{P\left(\nu_e\rightarrow \nu_\mu\right)}{\sin^2 2\theta}&\approx&   \frac{m_0^2}{4k^2}\sin^2(\Lambda t)
    +\frac{m_0^2 \Lambda^2}{16k^4}\left(\cos(\Lambda  t)-1\right)^2     \nonumber\\
    &\approx& \frac{m_0^2}{4k^2} \sin^2\left(\Lambda t\right) \, ,
    \label{probabilityGenNeutrino}
\end{eqnarray}
as the last term is negligible compared to the first one. The probability \eqref{probabilityGenNeutrino} resembles  the standard result for the two-neutrino oscillation problem \cite{kayser,mondal,fantini,petcov} when we identify  $\Lambda\equiv \Delta m^2/k\sim 10^{-10}$ eV, where the known neutrino mass difference is of the order  $\Delta m^2\sim 10^{-4}$ eV$^{2}$ \cite{Olive}. This value for $\Lambda$ is consistent with our initial assumption $\Lambda\ll m_0\ll k$; besides, it allows us to get that the oscillation in this model permit the neutrino to travel distances of the order
$L=c t\leq c\pi/\Lambda< 10^3$ km, which is consistent with the measurements \cite{Olive}.
 
Because of the above result \eqref{probabilityGenNeutrino},  one could infer that the neutrino T-SUSY oscillations are not produced by different mass states, but because of the appearing of the supersymmetry in time-domain due to the single neutrino non-constant mass.
Under this T-SUSY theory, the oscillation occurs due to the supersymmetric nature of the neutrino in time-domain. It is always one neutrino, with a unique mass, that it is oscillating between supersymmetric mass states.  

\section{Discussion}
When the mass of a relativistic quantum mechanical field is time-dependent, then
a supersymmetric in time-domain theory may be  constructed.
This T-SUSY is the time analogue of any relativistic supersymmetric quantum  theory. Along this work, we have presented  solutions for different time-dependent masses. A  remarkable outcome is that this  theory contains solutions that mimic the  light-like behavior studied for the bosonic T-SUSY theory.

 One of the main results extracted here is the possibility for probability oscillation between two states due to the temporal changes of field mass. These T-SUSY oscillations were applied to the two-neutrino oscillation problem, obtaining that the oscillation between flavour states can be explained by invoking only the time-dependence of a single neutrino mass (not different masses for different states).
Furthermore, the oscillation problem for more states (such as the three-neutrino oscillation) can be straightforwardly generalized from 
  Eq.~\eqref{wavefunccompoentmixed}. In this way, this   theory for spin fields introduces a different and interesting perspective to understand the neutrino oscillation problem under the new light of supersymmetry in time-domain.

The T-SUSY theory presented here may be its simplest, yet nontrivial,  version. The extension to other types of physical interactions is straightforward. For example, a T-SUSY version of the Dirac equation with Lorentz scalar potential \cite{coop2} may be constructed in an analogous way to what was presented here (in fact, the time-dependent mass can be understood as the interacting scalar potential field). In such case, the non-relativistic limit will coincide with  known theories \cite{Vladislavg,Vladislavg2,Vladislavg3,juliacen}.
Additionally, in the specific case of minimal coupling to electromagnetism \cite{coop2},
no T-SUSY are found in a simple way as presented here. 
On the other hand, extensions of this theory to spinor fields in curved spacetimes can also be proposed \cite{curvedsusydirac}, and it is currently under investigation. 

Finally, it is a matter of exploration to understand if the supersymmetric structures of this formalism may be closely linked to those appearing in time crystals \cite{wilck2}.

\end{document}